\documentclass[12pt]{article}

\newcommand{\newc}{\newcommand}
\newc{\gsim}{\lower.7ex\hbox{$\;\stackrel{\textstyle>}{\sim}\;$}}
\newc{\lsim}{\lower.7ex\hbox{$\;\stackrel{\textstyle<}{\sim}\;$}}
\newc{\gev}{\,{\rm GeV}}
\newc{\mev}{\,{\rm MeV}}
\newc{\ev}{\,{\rm eV}}
\newc{\kev}{\,{\rm keV}}
\newc{\tev}{\,{\rm TeV}}
\def\ln{\mathop{\rm ln}}

\newc{\mz}{M_Z}
\newc{\ms}{M_*}
\newc{\mpl}{M_{pl}}
\newc{\mw}{m_{\rm weak}}
\def\beq{\begin{equation}}
\def\eeq{\end{equation}}
\def\bea{\begin{eqnarray}}
\def\eea{\end{eqnarray}}
\newc{\ie}{{\it i.e.}}          \newc{\etal}{{\it et al.}}
\newc{\eg}{{\it e.g.}}          \newc{\etc}{{\it etc.}}
\newc{\cf}{{\it c.f.}}
\def\bar#1{\overline{#1}}

\def\inv{^{\raise.15ex\hbox{${\scriptscriptstyle -}$}\kern-.05em 1}}
\def\lbar{{\lower.35ex\hbox{$\mathchar'26$}\mkern-10mu\lambda}} 

\def\om#1#2{\omega^{#1}{}_{#2}}

\def\to{\rightarrow}

\let\<=\langle
\let\>=\rangle

\let\+=\uparrow

\let\Ga=\Gamma
\let\de=\delta

\let\si=\sigma

\let\om=\omega
\let\Om=\Omega

\begin{document}
\thispagestyle{empty}
\vspace*{.5cm}
\noindent
\hspace*{\fill}{\large CERN-TH/2002-014}\\
\vspace*{2.0cm}

\begin{center}
{\Large\bf Calculable Corrections to Brane Black Hole Decay I:}\\[.2cm]
{\Large\bf The Scalar Case}
\\[2.5cm]
{\large Panagiota Kanti and John March-Russell
}\\[.5cm]
{\it Theory Division, CERN, CH-1211 Geneva 23, Switzerland}
\\[.2cm]
(March, 2002)
\\[1.1cm]

{\bf Abstract}\end{center}
\noindent
In the context of brane-world theories, the production
cross-section for black holes may be greatly enhanced.  Such
black holes can in principle lead to detectable signals via
their Hawking evaporation to brane-localized modes.  We calculate,
in the semiclassical approximation, the leading corrections to 
the energy spectrum (the greybody factors) for decay into scalar
fields, as a function of the number of toroidally
compactified extra dimensions, and partial wave number. 
\newpage

\setcounter{page}{1}

\section{Introduction}

It is a remarkable fact that in brane-world theories the true scale,
$\ms$, of quantum gravity may be substantially lower than the
traditional Planck scale, $M_{pl}$, possibly approaching the TeV-scale,
and this radical departure from the standard picture is not excluded~\cite{ADD}
(for earlier works on brane theories see \cite{early}).
This observation has naturally excited a large amount of activity
investigating both the structure of these theories and their experimental
signals~\cite{activity}.  One
of the most striking consequences of lowering the Planck scale to the TeV
region is that the properties of small black holes are substantially
altered~\cite{admr}.
A black hole of given mass $M$ is now much lighter, larger, and colder than
a usual black hole of the same mass, provided only that the Schwarschild
radius $r_H$ of the black hole is smaller than the size of the extra
dimensions $r_H < R$.  In this limit, the black hole
is well-described as a $(4+n)$-dimensional black hole centered on the
brane, but extending out into the $n$ extra dimensions.
The horizon radius $r_H$ of such a black hole is~\cite{admr,mp},
\beq
r_H = {1\over\ms} \left(M\over\ms\right)^{1\over n+1}
\left(8 \Gamma((n+3)/2)\over (n+2) \pi^{(n+1)/2}\right)^{1/(n+1)} .
\label{eq:rh}
\eeq
(Note that, following common practice, we work in the approximation
that the brane tension itself does not strongly perturb the
$(4+n)$-dimensional black hole solutions.)

In particular it is likely (though not proven) that black holes are
much easier to produce, with
production cross-section at parton-parton
c.o.m. energies $\sqrt{s}$ close to the geometrical cross-section of
a black hole of mass $M=\sqrt{s}$~\cite{gt,dl} (for supporting
evidence see, \eg, \cite{support}, for claims to the contrary, see
\cite{voloshin})
\beq
\si_{\rm prod}(s) \simeq \pi r_H^2 = \frac{1}{\ms^2}\left(
\frac{M}{\ms} \left\{ \frac{8\Gamma\left( \frac{n+3}{2}\right) }{n+2} .
\right\}\right)^{2/(n+1)}
\label{eq:prod}
\eeq
If this is the case, then for $\ms\sim{\cal O}(\tev)$ there
are striking consequences for the high-energy interactions of 
cosmic rays~\cite{cosmic}, and, moreover, the LHC will become
a `black-hole factory'~\cite{gt,dl,bhphenom}.

After such black holes are produced they decay by the emission
of Hawking radiation.  It is expected that black holes
produced by a collision on our brane, \eg\  $p{\bar p} \to BH +X$
at a hadron collider, will decay mostly to particles on our brane~\cite{ehm},
and thus be indirectly observable via this characteristic Hawking
radiation.
This radiation
is usually described as `thermal' in character with a temperature
\beq
T_{BH}= {(1 + n) \over 4 \pi} {1\over r_H}
\label{eq:TBH}
\eeq
(in $G_N = k_B = c =\hbar=1$ units).  However, because
of the non-trivial metric in the region exterior to the horizon
there exists an effective potential barrier in this exterior region.
This potential barrier backscatters a part of the outgoing radiation
back into the black hole, the amount depending on the energy of the
radiation.  Thus the original blackbody radiation
is modified by a frequency-dependent filtering function,
$\si(\om)$, caused by the gravitational potential of the black hole.
The function, $\si(\om)$, is known as the `greybody factor'.
The black hole differential decay rate into particles of
energy $\om$ is then given by the Hawking formula~\cite{hawking}
\beq
\frac{dE(\om)}{dt} = \sum_{\ell,b} \sigma_{\ell,b}(\om) {\om  \over
\exp\left(\om/T_{BH}\right) \mp 1} \frac{d^{n+3}k}{(2\pi)^{n+3}}
\label{eq:greybody}
\eeq
where $\ell$ labels the angular momentum quantum number,
$b$ labels any other quantum numbers of the emitted
particle, as well as the particle type, and in the phase-space
integral $|k| = \om$ for a massless particle.\footnote{An alternate
form for the decay rate that is sometimes useful in the massless particle case
involves the absorption probability $|{\cal A}(\om)|^2$ whose relation to $\si(\om)$
is given in Eq.(\ref{grey}): $dE(\om)/dt = \sum_{\ell,m,b} |{\cal A}(\om)|^2 \om d\om/
[2\pi(\exp\left(\om/T_{BH}\right) \mp 1)]$.  Here $m$ is the azimuthal quantum number.}

Greybody factors are important theoretically because they encode
information on the near horizon structure of black holes.  Indeed
one of the most exciting features of BH production at the LHC would
be the opportunity to investigate Hawking emission in detail for
clues as to how the
infamous apparent violation of the laws of quantum mechanics by
black hole evaporation (the `information paradox' of black
holes~\cite{preskill}) is resolved -- if it is, that is!

Greybody factors can
be important experimentally because they modify the spectrum in the
region where most particles are produced thus altering the
characteristic spectrum by which we hope to identify a `BH event'.
In particular the functional dependence of $\si_b(\om)$ on the energy
$\om$ depends on the spin of the emitted particle, and on whether it
is brane-localised or free to propagate in the bulk of the extra
dimensions.

One can compute the greybody factor by first computing
the absorption cross section for the appropriate type of particle
incident on the background metric that describes the brane black hole.
This is because the greybody factor in the Hawking formula
for the emission rate of a
given type of outgoing particle, $b$, at energy $\om$ equals the
absorption cross section $\si_{b,abs}(\om)$ for a particle of type
$b$ incoming at energy $\om$.  In fact it is this property which
implies the greybody factors do not invalidate the thermal nature of the
black hole.  Since the outgoing transmission and incoming absorption
coefficients are equal to one another, equilibrium still occurs
if the black hole is placed in a heat bath.

The semiclassical calculation of Hawking emission is only reliable when
the energy of the emitted particle is small compared to the black
hole mass $\om \ll M$, since only in this case is it correct
to neglect the back reaction of the metric during the emission process.
This in turn requires that the Hawking temperature $T_{BH}\ll M$, which
is equivalent to demanding that the black hole mass $M\gg \ms$, as can
be seen from Eqs.(\ref{eq:TBH}) and (\ref{eq:rh}).  Inevitably this condition
breaks down during the final stages of the decay process, but for those
black holes of initial mass larger than $\ms$ most of the evaporation
process is well-described by the semi-classical calculation.

In this paper we calculate the greybody factor for the simplest
case of scalar particles, both free to propagate in the bulk,
and localized on the brane (\eg, the Higgs).  Section~2 contains
a discussion of the metric used in calculating the greybody factors
and the quality of the approximations used.
To orient the reader through our analysis, Section~3 presents the
calculation of
bulk scalar emission in $(4+n)$-dimensions, in the simple case of S-wave
emission.  Section~4 performs the calculation of the greybody factor
for bulk scalar emission for arbitrary
partial wave.  The primary result of this section, the general formula
for the greybody factor, is given in Eq.(\ref{siresult}).
Section~5 contains the calculation of brane-localized scalar emission,
where the brane is embedded in a $(4+n)$-dimensional bulk.
The primary results of this section are given in Eqs.(\ref{sibrane}) and
(\ref{finalresult}).
We conclude in Section~6.
For ease of use, the results of our calculations for the greybody factors,
for the most important ($\ell =0,1,2$) angular momentum modes, and for
the $n=2,4,6$ extra dimensions are collected in Tables~1 and 2.


\section{Black hole metrics and greybody factors}

Let $V=(2\pi R)^n$ be the volume of the $n$ extra
dimensions, which are here taken to be of common radius $R$.  Then
Gauss' Law relates the 4d Planck mass $\mpl$ to the new fundamental
scale of gravity by $\mpl^2 = V_n \ms^{2+n}$.  A black hole of mass
$M\ll \mpl (\mpl/\ms)^{(2+n)/n}\sim 10^{15+32/n}\gev$ is, for
distances $r\ll R$, well approximated~\cite{admr} by
the $(4+n)$-dimensional Schwarzschild black hole with line-element
\beq
ds^2 = - h(r)dt^2 +h(r)^{-1} dr^2 + r^2 d\Omega_{2+n}^2,
\label{eq:bhmetric}
\eeq
where
\beq
h(r) = 1-\biggl(\frac{r_H}{r}\biggr)^{n+1}\,,
\label{h-fun}
\eeq
with the horizon radius, $r_H$, given in Eq.(\ref{eq:rh}).  In
Eq.(\ref{eq:bhmetric}) the angular part is
\beq
d\Omega_{2+n}^2=d\theta^2_{n+1} + \sin^2\theta_{n+1} \,\biggl(d\theta_n^2 +
\sin^2\theta_n\,\Bigl(\,... + \sin^2\theta_2\,(d\theta_1^2 + \sin^2 \theta_1
\,d\varphi^2)\,...\,\Bigr)\biggr)\,,
\eeq
with $0 < \varphi < 2\pi$ and $0<\theta_i<\pi$, for $i=1,...,n+1$.
However, because of the compactification of the extra dimensions, the metric
for this black hole at distances $r\gg R$ goes over to that of the usual
4-dimensional Schwarzschild solution
\beq
ds^2 = - \left(1 - \frac{2M}{\mpl^2 r}\right) dt^2 +
\left(1 - \frac{2M}{\mpl^2 r}\right)^{-1} dr^2 + r^2 d\Omega_{2}^2\, .
\label{eq:4dbhmetric}
\eeq

The matching of the two expressions, Eqs.(\ref{eq:bhmetric}) and
(\ref{eq:4dbhmetric}) at $r\simeq R$ is of course just an approximation
to the exact metric for such a black hole.   An exact expression
is not necessary to derive the form of the greybody factor to
a very good approximation for the energies of interest.

To understand this consider the case where a scalar field propagates in the
full $(4+n)$-dimensional Schwarzschild black hole background.
The scalar wave equation $g^{IJ} \phi_{,I;J}=0$ in this background
is separable if we make the ansatz
\beq
\phi(t,r,\theta_i,\varphi)=
e^{-i\om t}\,R_{\om \ell}(r)\,{\tilde Y}_\ell(\Om)\, ,
\eeq
where ${\tilde Y}_\ell(\Om)$ is the $(3+n)$-spatial-dimensional
generalisation of the usual spherical harmonic functions depending on
the angular coordinates~\cite{spherical}.  Upon substituting this ansatz,
the scalar
wave equation implies a second-order differential equation for the
radial wavefunction $R_{\om \ell}(r)$:
\beq
\frac{h(r)}{r^{n+2}}\,\frac{d \,}{dr}\,
\biggl[\,h(r)\,r^{n+2}\,\frac{d R}{dr}\,\biggr] +
\biggl[\,\om^2 - \frac{h(r)}{r^2}\,\ell\,(\ell+n+1)\,\biggr] R =0 \, .
\label{scalareqn}
\eeq
The greybody factor for BH decay into scalars will be found from the
solutions to this equation.

We can transform this equation to a more convenient form by
defining a new (``tortoise") radial coordinate $y$ by
\beq
y=\frac{\ln h(r)}{r_H^{n+1}\,(n+1)} \,\,\Rightarrow \,
\frac{dy}{dr} = \frac{1}{ h(r)\,r^{n+2} }
\label{ycoord}
\eeq
in terms of which the radial equation becomes
\beq
\left(\frac{d^2 \,}{dy^2} + r^{2n+4}\left[\omega^2 -\frac{\ell(\ell+n+1)
h(r)}{r^2}\right] \right) R(y) = 0\,.
\label{yeqn}
\eeq
Since the
coordinate position $r_H$ of the horizon is defined by the (largest)
solution to $h(r_H)=0$, the horizon in terms of $y$ is at $y\to -\infty$,
as can been seen from Eq.(\ref{ycoord}).  Eq.(\ref{yeqn}) is analogous
to the Schrodinger equation for a particle in an effective potential.

Non-trivial backscattering occurs in this metric
when the 2nd of the two terms in the []-parentheses in
Eq.(\ref{yeqn}) is comparable or larger than the 1st.  For black holes
at the LHC the typical energies of emitted radiation we are interested in
range from $\om_{max}\sim T_{BH} = {(1 + n)/ 4 \pi  r_H}
\sim ({\rm few}) 100\gev$ to a minimum
of $\sim({\rm few})\gev$ (this minimum might be set by the energy threshold of
the detectors -- the precise value will not matter).  For the range of
parameters, $n$, and $\ms$, of interest, this gives a range of $\om$ from
$\om_{max}\lsim 1/2 r_H$ to $\om_{min} \gsim 1/200 r_H$.  Thus, inspecting
the potential terms of Eq.(\ref{yeqn}) we see that the potential is only
large enough to lead to backscattering over the range of distances $r$
from ${\cal O}(1) r_H$ to ${\cal O}(100) r_H$.

A similar analysis can be performed for the 4d asymptotic metric of
Eq.(\ref{eq:4dbhmetric}).  In this case one finds that significant
backscattering of the quanta of interest would only occur when
distances were of order $r\sim M/\mpl^2 \ll r_H \ll R$.  But at
such distances the 4d asymptotic form of the metric is not
applicable, being replaced
by the $(4+n)$-dimensional metric used above.  Instead what happens is
that the change-over from the $(4+n)$-dimensional to 4-dimensional
regime, at distances of order $r\sim R$, only significantly backscatters
quanta of energy $10^{-3}\ev$ or less.
In other words, the backscattering due to the effect of the compactification
of the extra dimensions only affects very low energy quanta which are
experimentally irrelevant.  Even more so this applies to the change
in backscattering at distances $r\sim R$ due to the difference between
the approximate form of the black hole metric, Eqs.~(\ref{eq:bhmetric}) and
(\ref{eq:4dbhmetric}), and the exact black hole solution.

Overall, the lesson is that upto corrections that only apply for very
low energy, the greybody factors for brane black holes may be calculated
using purely the $(4+n)$-dimensional regime of the black hole metric.
In the remainder of this paper we will follow this procedure.  Thus
our task is now to return to the radial equation Eq.(\ref{scalareqn})
or (\ref{yeqn}) and compute the greybody factor.

In particular, via the equivalence of $\si(\om)$ to the absorption
cross section, the greybody factor can be computed by
first evaluating the absorption probability, $|{\cal A}(\om)|^2$,
from the ratio of the in-going flux at the future horizon to
the incoming flux from past infinity (with the boundary condition
that there is no outgoing flux at the horizon), and then using
the generalised $(4+n)$-dimensional optical theorem relation~\cite{gkt}
\beq
\si_\ell(\om) = \frac{2^{n}\pi^{(n+1)/2}\,\Ga[(n+1)/2]}{ n!\, \om^{n+2}}\,
\frac{(2\ell+n+1)\,(\ell+n)!}{\ell !}\, |{\cal A}|^2
\label{grey}
\eeq
between the absorption cross section $\si_\ell(\om)$ and the absorption
probability $|{\cal A}|^2$ for the $\ell$'th partial wave. This formula includes
a summation over the multiplicity of individual `azimuthal' components
for each given partial wave $\ell$ (i.e. $2\ell+1$ in 3 spatial dimensions,
$(\ell+1)^2$ in 4 spatial dimensions, etc.). A simple summation is
appropriate because we deal with a non-rotating black hole.

The radial equation, however, is not in general exactly soluble,
therefore, we will employ an approximation method based on splitting
the radial domain into `near-horizon' (NH) and
`far-field' (FF) regions.  The solutions (satisfying appropriate boundary
conditions) in these two regions will be computed and then matched
in a transition region to find the complete solution.  This procedure
leads to an expression for $|{\cal A}_\ell(\om)|^2$ correct in leading
order in an expansion in the dimensionless quantity $\om r_H$.


\section{Bulk scalar emission: S-wave example}

Because the radius and temperature of the $(4+n)$-dimensional
black hole are always comparable, Eq.(\ref{eq:TBH}), the dominant
scalar decay mode is that into the S-wave, $\ell=0$.
We will first solve the problem in this case, which will also serve as an
illuminating example for the full case studied in Sections~4 and 5.

We will first compute the solution in the `near-horizon' (NH) region.
The radial equation for the S-wave, in terms of the $y$-coordinate
defined in Eq. (\ref{ycoord}), becomes
\beq
\biggl(\frac{d^2 \,}{dy^2} + \omega^2 r^{2(n+2)}\biggr) R(y) = 0\,.
\label{yeqnS}
\eeq
By expanding close to the horizon
$r = r_H + \de r$ ($\de r \ll r_H$), we obtain
$\de r \simeq r_H \exp\left( (n+1) r_H^{n+1} y\right)$ as $y\to -\infty$.
Thus the radial equation Eq.(\ref{yeqnS}) becomes in the near-horizon limit
\beq
\biggl(\frac{d^2 \,}{dy^2} + \omega^2 r_H^{2(n+2)}\biggr) R(y) = 0\,,
\eeq
up to exponentially small corrections in $y$.  The general
near-horizon solution is therefore
\beq
R_{NH}(y) = A_+ \exp(ir_H^{n+2} \om y) + A_- \exp(-ir_H^{n+2} \om y) .
\label{NHscalar}
\eeq
In order to calculate the greybody factor, we must impose the boundary condition
that near the horizon the solution is pure in-going. Therefore, we need to
set $A_+=0$.

We now turn to the far-field region which is defined by $r \gg r_H$. In this
limit, $h(r) \simeq 1$ and, by setting $R(r)=f(r)/r^{(n+1)/2}$,
Eq.(\ref{scalareqn}) can be rewritten as
\beq
\frac{d^2 f}{dr^2} +\frac{1}{r}\,\frac{d f}{dr} +
\Bigl[\,\om^2 - \frac{(n+1)^2}{4r^2}\,\Bigl] f=0\,,
\eeq
which has the form of a Bessel differential equation \cite{abram}. The
general solution for the radial function $R(r)$ is therefore given by
\beq
R_{FF}(r) = \frac{B_+}{r^{(n+1)/2}}\,J_{(n+1)/2}(\om r) +
\frac{B_-}{r^{(n+1)/2}}\,Y_{(n+1)/2}(\om r)\,,
\label{FFscalar}
\eeq
where $J_{(n+1)/2}(\om r)$ and $Y_{(n+1)/2}(\om r)$ are the Bessel functions
of the first and second kind, respectively.

As we will soon see, the two coefficients $B_+$ and $B_-$, and more
specifically their ratio, will help us define the greybody factor.
To compute this ratio, we need to match the far-field solution,
Eq.(\ref{FFscalar}), on to the near-horizon solution, Eq.(\ref{NHscalar}),
in the intermediate region. To this end, we expand the near-horizon solution,
in the regime $\om r\ll 1$ and $r\gg r_H$, leading to
\footnote{This expansion allows $\om \lsim 1/r_H$, the typical emitted energy,
when $M\gg \ms$, as is anyway required for the reliability of the
semiclassical approach.}
\beq
R_{NH}(r) \simeq A_- \exp\left\{ i\frac{\om r}{n+1}
\left(\frac{r_H}{r}\right)^{n+2} \right\} \simeq
A_-\left\{ 1 + i\frac{\om r}{n+1} \left(\frac{r_H}{r}\right)^{n+2} \right\}\,.
\label{NHlarger}
\eeq
We also expand the far-field solution, Eq.(\ref{FFscalar}), in the same
regime $\om r \ll 1$, which gives
\beq
R_{FF}(r) \simeq \frac{B_+}{\Ga(\frac{n+3}{2})} \left(\frac{\om}{2}\right)^{(n+1)/2}
- \frac{B_-}{r^{n+1}} \left(\frac{2}{\om}\right)^{(n+1)/2}
\frac{\Ga(\frac{n+1}{2})}{\pi}\,.
\eeq
Matching the above expression with Eq.(\ref{NHlarger}), we find the result
\beq
\frac{B_+}{B_-} = i \frac{\Ga(\frac{n+3}{2})^2\,2^{n+2}}
{\pi\,(\om r_H)^{n+2}}\,.
\eeq

The reflection coefficient ${\cal R}$ for scattering in the gravitational
potential of the black hole Eq.(\ref{eq:bhmetric}) is defined as the ratio
of the outgoing and incoming amplitude at infinity. To compute this,
we expand the far-field solution Eq.(\ref{FFscalar}), in the limit
$\om r \rightarrow \infty$, which yields
\beq
R_{FF}(r) \simeq \frac{(B_+ - i B_-)}{\sqrt{2\pi \om r^{n+2}}}\,
e^{i\Bigl(\om r -(n+2)\pi/4\Bigr)} +
\frac{(B_+ + i B_-)}{\sqrt{2\pi \om r^{n+2}}}\,
e^{-i\Bigl(\om r -(n+2)\pi/4\Bigr)}\,,
\eeq
and which, in turn, leads to the following expression for the
reflection coefficient
\beq
{\cal R} = {{\rm outgoing~amplitude}\over {\rm incoming~amplitude}}
= \frac{B_+ -iB_-}{B_+ + iB_-}\,,
\label{abs-coef}
\eeq
up to a purely imaginary phase that will drop out when the magnitude of
${\cal R}$ will be computed. The absorption probability is then defined as
\beq
|{\cal A}|^2 = (1 - |{\cal R}|^2) \simeq  \frac{\pi\,(\om r_H)^{n+2}}
{2^n\,\Ga(\frac{n+3}{2})^2}\,,
\label{eq:simpleres}
\eeq
where, in the final expression, we have expanded to leading order in
$(\om r_H)$.
This is the final result for the S-wave greybody factor for scalars
in a $(4+n)$-dimensional Schwarzschild black hole background. We may easily
check that, for $n=1$, we correctly reproduce the result
\beq
|{\cal A}|^2 = \frac{\pi}{2}\,\om^2\,r_H^3 = \frac{\om^3}{4 \pi}\,A_H\,,
\eeq
where $A_H=2 \pi^2 r_H^3$ is the area of the horizon, presented in
Ref.~\cite{DasMathur}.


\section{Bulk Scalar emission for $\ell\ge 0$}

We will now generalise the above analysis in the case where the scalar modes
are not spherically symmetric, $\ell \neq 0$.  For readers more
interested in our final results, rather than the techniques used to solve
the problem, we suggest jumping to Eq.(\ref{siresult}) and
following discussion.

We will start with
the derivation of the solution in the near-horizon zone. Starting from
Eq.(\ref{scalareqn}), and making a change of variable, we may write
the scalar field equation in the form (here, we adopt the method of
Ref.~\cite{maldacena})
\beq
h\,(1-h)\,\frac{d^2R}{dh^2} + (1-h)\,\frac{d R}{dh} +
\biggl[\,\frac{(\om r)^2}{(n+1)^2 h (1-h)} -
\frac{\ell\,(\ell + 1 +n)}{(n+1)^2 (1-h)}\,\biggr] R=0\,.
\label{NH-ell-1}
\eeq
Near the horizon, $r \simeq r_H$ and the quantity $(\om r)^2$ can be set equal
to $(\om r_H)^2$. Then, by using the redefinition
$R(h)=h^\alpha (1-h)^\beta F(h)$, the above equation takes the form of
the hypergeometric equation
\beq
h\,(1-h)\,\frac{d^2 F}{dh^2} + [c-(1+a+b)\,h]\,\frac{d F}{dh} -ab\,F=0\,,
\label{hyper}
\eeq
with $a=b=\alpha + \beta$ and $c=1+2 \alpha$, where
\beq
\alpha_{\pm} = \pm \frac{i \om r_H}{n+1}\,, \qquad
\beta_{\pm} =\frac{1}{2} \pm \frac{1}{n+1}\,
\sqrt{\Bigl(l +\frac{n+1}{2}\Bigr)^2 -(\om r_H)^2}\,.
\label{al-be}
\eeq
Equation (\ref{hyper}) has as a solution the hypergeometric function $F(a,b,c;h)$.
The criterion for the convergence of the hypergeometric function demands that
${\cal R}(c-a-b)>0$, which forces us to choose $\beta=\beta_-$. Then, the general
solution of Eq.(\ref{NH-ell-1}) may be written as~\cite{abram}
\beq
R_{NH}(h)=A_- h^{\alpha_{\pm}}\,(1-h)^\beta\,F(a,b,c;h) +
A_+\,h^{-\alpha_{\pm}}\,(1-h)^\beta\,F(a-c+1,b-c+1,2-c;h)\,.
\label{NH-gen}
\eeq
Expanding the above solution in the limit $r \rightarrow r_H$, or
$h \rightarrow 0$,
and choosing $\alpha=\alpha_-$, we obtain the result
\beq
R_{NH} \simeq \Bigl(\frac{r_H}{r}\Bigr)^{\beta (n+1)}\,\biggl[\,A_-\,
\exp\Bigl(-i \om r_H^{n+2} y\Bigr) +
A_+\,\exp\Bigl(i \om r_H^{n+2} y\Bigr)\,\biggr]\,,
\label{NH-nh}
\eeq
which again imposes the condition\footnote{Note that the choice
$\alpha=\alpha_+$ would have led again to Eq.(\ref{NH-nh}) with
$A_- \leftrightarrow A_+$, and therefore to the choice $A_-=0$.
As both values of $\alpha$ appear in the general solution Eq.(\ref{NH-gen}),
it is only a matter of choice which one of the two terms will be
associated with the incoming mode.} $A_+=0$.

The derivation of the far-field-zone solution closely follows the analysis
performed in the case where $\ell=0$. The same redefinition of the radial
function leads again to a Bessel equation whose general solution is now
given by
\beq
R_{FF}(r) = \frac{B_+}{r^{(n+1)/2}}\,J_{\ell + (n+1)/2}(\om r) +
\frac{B_-}{r^{(n+1)/2}}\,Y_{\ell + (n+1)/2}(\om r)\,.
\label{FFscalar-ell}
\eeq

Both solutions, near-horizon and far-field, need to be ``stretched" and
matched in the intermediate region. Expanding first Eq.(\ref{FFscalar-ell}),
in the limit $\om r\ll 1$, gives
\beq
R_{FF}(r) \simeq \frac{B_+ \,r^\ell}{\Ga(\ell + \frac{n+3}{2})}
\left(\frac{\om}{2}\right)^{\ell + (n+1)/2}
- \frac{B_-}{r^{\ell+n+1}} \left(\frac{2}{\om}\right)^{\ell+(n+1)/2}
\frac{\Ga(\ell+\frac{n+1}{2})}{\pi}\,.
\label{FFsmall-ell}
\eeq

The near-horizon solution needs to be ``shifted" first and expressed in terms
of $1-h$, before being expanded in the limit $r \gg r_H$. By using a standard
formula \cite{abram}, we write
\bea
\hspace*{-0.4cm}
R_{NH}(h)&=&A_-\,h^{\alpha}\,\biggl[\,(1-h)^\beta \,\frac{\Gamma(1+2\alpha)\,
\Gamma(1-2\beta)}{\Gamma(1+\alpha-\beta)^2}\,F(a,b,a+b-c+1;1-h) \nonumber\\[2mm]
&+&
(1-h)^{1-\beta} \,\frac{\Gamma(1+2\alpha)\,
\Gamma(2\beta-1)}{\Gamma(\alpha+\beta)^2}\,F(c-a,c-b,c-a-b+1;1-h)\,\biggr].
\eea
We can now expand the above expression in the limit $h \rightarrow 1$
and take
\beq
R_{NH}(h) \simeq A_-\,\Bigl(\frac{r}{r_H}\Bigr)^\ell\,\frac{\Gamma(1+2\alpha)\,
\Gamma(1-2\beta)}{\Gamma(1+\alpha-\beta)^2}
+ A_-\,\Bigl(\frac{r_H}{r}\Bigr)^{\ell + n +1}\,\frac{\Gamma(1+2\alpha)\,
\Gamma(2\beta-1)}{\Gamma(\alpha+\beta)^2}\,.
\label{NHlarger-ell}
\eeq

Matching the two solutions Eq.(\ref{FFsmall-ell}) and Eq.(\ref{NHlarger-ell}), we
obtain the ratio
\beq
\frac{B_+}{B_-} = -\biggl(\frac{2}{\om r_H}\biggl)^{2\ell+n+1}\,
\frac{\Ga(\ell +\frac{n+1}{2})^2\,\Bigl(\ell +\frac{n+1}{2}\Bigl)\,
\Ga(1-2\beta)\,\Ga(\alpha+\beta)^2}{\pi\,\Ga(1+\alpha-\beta)^2\,
\Ga(2\beta-1)}\,.
\eeq
The definition of the reflection coefficient ${\cal R}$ is still given
by Eq.(\ref{abs-coef}). In turn, the absorption probability can
be written, in terms of $B=B_+/B_-$, as
\beq
|{\cal A}|^2 = (1 - |{\cal R}|^2) = \frac{2i\,(B^*-B)}{B B^* +
i(B^*-B) +1}\,.
\label{sigma}
\eeq
Due to the fact that the argument of the Gamma functions appearing
in the expression of ${\cal R}$ are non-trivial complex numbers,
we can not write the absorption coefficient in a simple way in the
general case. However, the above expression can be further simplified
in the limit $\om r_H \ll 1$, in which case $B B^* \gg i(B^*-B) \gg 1$,
and, therefore, $|{\cal A}|^2$ may be written as
\beq
|{\cal A}|^2= \frac{4 \pi^2}{2^{4 \ell/(n+1)}}\,
\biggl(\frac{\om r_H}{2}\biggl)^{2\ell+n+2}\,
\frac{\Ga(1+\frac{\ell}{n+1})^2}
{\Ga(\frac{1}{2}+\frac{\ell}{n+1})^2\,\Ga(\ell +\frac{n+3}{2})^2}\, .
\label{siresult}
\eeq
This is our major result for the case of bulk scalar fields.
Eq.(\ref{siresult}) nicely displays the leading functional
dependence of the
greybody factor on $\om r_H$ for varying partial wave, $\ell$,
and number of extra dimensions, $n$.
For the case of $\ell=0$, this may be further
evaluated to give
\beq
|{\cal A}|^2_{\ell=0} = \biggl(\frac{\om r_H}{2}\biggl)^{n+2}\,
\frac{4\pi}{\Ga[(n+3)/2]^2}\,, 
\label{siresultzero}
\eeq
in complete agreement with the result, Eq.(\ref{eq:simpleres}),
of our earlier analysis in the case $\ell=0$.
In fact, for an $s$-wave massless bulk scalar, 
the absorption probability $|{\cal A}|^2$ has the exact form that allows
the greybody factor $\sigma (\omega)$ to reduce to the horizon area
of the black hole in agreement with previous work~\cite{gdm}.
The numerical value of the results for the cases $\ell=0,1,2$
for $n=2,4$ and $6$ are shown in Table~1.

If we fix the number
of extra dimensions and vary only the angular momentum number, 
the absorption probability decreases as $\ell$ increases. This
decrease is caused by the fact that the power of the expansion parameter
$(\omega r_H)^{2\ell+n+2}$, in the leading term, increases with
$\ell$. Since $(\omega r_H) \ll 1$, this means that $|{\cal A}|^2$ 
becomes more and more suppressed as $\ell$ increases. The numerical
coefficient in front of the leading term also decreases with $\ell$
(see Table 1).
The same behaviour is observed if we fix instead $\ell$ and vary $n$.


%
\begin{center}
$\begin{array}{|c||c||l|} \hline \hline 
\multicolumn{3}{|c|}{\rule[-3mm]{0mm}{8mm} 
{\bf Table~1: \rm \,\,Absorption\,\,probabilities\,\,for\,\,a\,\,
(4+n)\,\,bulk\,\,scalar\,\,field}} \\ \hline\hline
{\rule[-3mm]{0mm}{8mm}
\hspace*{0.8cm}{\bf n=2} \hspace*{0.8cm}} & \hspace*{0.8cm} \ell=0 \hspace*{0.8cm}
& \hspace*{0.8cm} 
|{\cal A}|^2 \simeq \frac{4}{9}\,(\om r_H)^4 +\ldots \hspace*{0.8cm}
\\[3mm]
 & \ell=1 &
\hspace*{0.8cm}  |{\cal A}|^2 \simeq \frac{2^{2/3}}{(15)^2}\,
\frac{\Ga(4/3)^2}{\Ga(5/6)^2}\,\pi\,(\om r_H)^6 + \ldots \\[3mm]
 & \ell=2 &
\hspace*{0.8cm} |{\cal A}|^2\simeq \frac{2^{-2/3}}{(105)^2}\,
\frac{\Ga(5/3)^2}{\Ga(7/6)^2}\,\pi\,\,(\om r_H)^8 + \ldots
\\[3mm] \hline
{\rule[-3mm]{0mm}{8mm} {\bf n=4}  }& \ell=0 &
\hspace*{0.8cm} |{\cal A}|^2 \simeq \frac{4}{(15)^2}\,(\om r_H)^6 + \ldots
\\[3mm]
 & \ell=1 &
\hspace*{0.8cm} |{\cal A}|^2\simeq \frac{2^{6/5}}{(105)^2}\,
\frac{\Ga(6/5)^2}{\Ga(7/10)^2}\,\pi\,\,(\om r_H)^8 + \dots
\\[3mm]
 & \ell=2 &
\hspace*{0.8cm} |{\cal A}|^2 \simeq \frac{2^{2/5}}{(105)^2\,3^4}\,
\frac{\Ga(7/5)^2}{\Ga(9/10)^2}\,\pi\,\,(\om r_H)^{10} + \ldots
\\[3mm] \hline 
{\rule[-3mm]{0mm}{8mm} {\bf n=6} } & \ell=0 &
\hspace*{0.8cm} |{\cal A}|^2 \simeq \frac{4}{(105)^2}\,(\om r_H)^8 + \ldots
\\[3mm]
 & \ell=1 &
\hspace*{0.8cm} |{\cal A}|^2 \simeq \frac{2\,2^{3/7}}{(105)^2\,3^4}\,
\frac{\Ga(8/7)^2}{\Ga(9/14)^2}\,\pi\,\,(\om r_H)^{10} + \ldots
\\[3mm]
 & \ell=2 &
\hspace*{0.8cm} |{\cal A}|^2 \simeq \frac{2^{6/7}}{(1155)^2\,3^4}\,
\frac{\Ga(9/7)^2}{\Ga(11/14)^2}\,\pi\,\,(\om r_H)^{12} + \ldots
\\[3mm] \hline \hline

\end{array}$
\end{center}


\section{Brane-localized scalar emission for $\ell\ge 0$}

We now turn to the study of the case where the scalar field is confined
on a four-dimensional brane embedded in a $(4+d)$-dimensional
Schwarzschild spacetime. The scalar field propagates in a four-dimensional
background whose metric tensor is given by the induced metric at the
location of the brane. The induced metric follows from the
$(4+d)$-dimensional one by fixing the values of
the extra angular coordinates: $\theta_n=\pi/2$ for $n \ge 2$, and it
may be written as
\beq
ds^2 = - h(r)dt^2 +h(r)^{-1} dr^2 + r^2\,(d\theta^2 +
\sin^2\theta\,d\varphi^2),
\label{ind-metric}
\eeq
where $h(r)$ is still given by Eq.(\ref{h-fun}).
The scalar field equation may be separated in the same way
\beq
\phi(t,r,\theta,\varphi)= e^{-i\om t}\,R_{\om \ell}(r)\,Y_\ell(\Om)\,,
\eeq
where $Y_\ell(\Om)$ are now the usual 3-dimensional spherical harmonic
functions. The above ansatz allows us to write the equation for the
radial part as
\beq
\frac{h(r)}{r^2}\,\frac{d \,}{dr}\,\biggl[\,h(r)\,r^2\,\frac{d R}{dr}\,\biggr] +
\biggl[\,\om^2 - \frac{h(r)}{r^2}\,\ell\,(\ell+1)\,\biggr] R =0\,.
\label{scaleqn-p}
\eeq
The presence of the metric function $h(r)$ makes once again the derivation
of the general solution extremely difficult. We will follow the same method
as in the previous section and compute the solution in the two radial domains,
near-horizon and far-field, which will then be ``stretched" and matched
in the intermediate region.

Having become familiar with the analysis, we will proceed to derive
directly the solution in the general case $\ell\ge 0$. We start with
the solution in the `near-horizon' (NH) region. In terms of $h$, the
radial differential equation now takes the form
\beq
h\,(1-h)\,\frac{d^2R}{dh^2} + \biggl[\,1 - \frac{(2n+1)}{(n+1)}\,h\,\biggr]\,
\frac{d R}{dh} + \biggl[\,\frac{(\om r_H)^2}{(n+1)^2 h (1-h)} -
\frac{\ell\,(\ell + 1)}{(n+1)^2 (1-h)}\,\biggr] R=0\,.
\label{NH-ell-p}
\eeq\\
By using the same redefinition $R(h)=h^\alpha (1-h)^\beta F(h)$, the above
equation assumes the standard form of a hypergeometric equation with
indices $a=\alpha + \beta + \frac{n}{(n+1)}$, $b=\alpha + \beta$ and
$c=1+2 \alpha$, where
\beq
\alpha_{\pm} = \pm \frac{i \om r_H}{n+1}\,, \qquad
\beta_{\pm} =\frac{1}{2(n+1)}\,\biggl(1 \pm
\sqrt{(2l +1)^2 -4(\om r_H)^2}\biggl)\,.
\label{al-be-p}
\eeq
The criterion for the convergence of the hypergeometric function,
${\cal R}(c-a-b)>0$, demands again $\beta=\beta_-$. The general
solution of Eq.(\ref{NH-ell-p}) has again the form of Eq.(\ref{NH-gen}).
Expanding near the horizon and imposing the condition that only
incoming waves exist near $r \simeq r_H$, we find that $A_+=0$
for $\alpha=\alpha_-$. At this point, we can also ``shift" the
solution and write it in terms of $1-h$, in the same way as in the
previous section. If we finally expand for $r \gg r_H$, or
equivalently $h \rightarrow 1$, we obtain the solution
\bea
&~& \hspace*{-6cm} R_{NH}(h) \simeq
A_-\,\Bigl(\frac{r}{r_H}\Bigr)^\ell\,\frac{\Gamma(1+2\alpha)\,
\Gamma(1-2\beta-\frac{n}{n+1})}{\Gamma(1+\alpha-\beta-\frac{n}{n+1})\,
\Gamma(1+\alpha-\beta)} \nonumber \\[3mm]
\hspace*{4cm} &~& +\,\,
A_-\,\Bigl(\frac{r_H}{r}\Bigr)^{\ell+1}\,\frac{\Gamma(1+2\alpha)\,
\Gamma(2\beta+\frac{n}{n+1}-1)}{\Gamma(\alpha+\beta+\frac{n}{n+1})\,
\Gamma(\alpha+\beta)}\,.
\label{NHlarger-ell-p}
\eea

The far-field-zone solution can be easily found to be given in terms
of the Bessel functions $J_{\ell + 1/2}(\om r)$ and
$Y_{\ell + 1/2}(\om r)$. Expanding the general solution
in the limit $\om r\ll 1$, finally gives
\beq
R_{FF}(r) \simeq \frac{B_+ \,r^\ell}{\Ga(\ell + \frac{3}{2})}
\left(\frac{\om}{2}\right)^{\ell + 1/2}
- \frac{B_-}{r^{\ell+1}} \left(\frac{2}{\om}\right)^{\ell+1/2}
\frac{\Ga(\ell+\frac{1}{2})}{\pi}\,.
\label{FFsmall-ell-p}
\eeq
Matching the two asymptotic solutions, we obtain the ratio
\beq
\frac{B_+}{B_-} = -\biggl(\frac{2}{\om r_H}\biggl)^{2\ell+1}\,
\frac{\Ga(\ell +\frac{1}{2})^2\,\Bigl(\ell +\frac{1}{2}\Bigl)\,
\Ga(1-2\beta-\frac{n}{n+1})\,\Ga(\alpha+\beta)\,
\Ga(\alpha+\beta+\frac{n}{n+1})}{\pi\,\Ga(1+\alpha-\beta)\,
\Ga(1+\alpha-\beta-\frac{n}{n+1})\,\Ga(2\beta +\frac{n}{n+1}-1)}\,,
\label{B-p}
\eeq
which can be used to determine the absorption coefficient according
to Eq.(\ref{sigma}). We may, however, obtain a simplified expression, in
the limit $\om r_H \ll 1$, which reads
\beq
|{\cal A}|^2= \frac{16 \pi}{(n+1)^2}\,
\biggl(\frac{\om r_H}{2}\biggl)^{2\ell+2}\,
\frac{\Ga(\frac{\ell+1}{n+1})^2\,\Ga(1+\frac{\ell}{n+1})^2}
{\Ga(\frac{1}{2}+\ell)^2\,\Ga(1 +\frac{2\ell+1}{n+1})^2}\,.
\label{sibrane}
\eeq
The expression of the absorption coefficient $|{\cal A}|^2$ for the
values $n=2,4$ and 6 of the number of extra dimensions, and $\ell=0,1,2$
of the angular momentum number, are shown in Table~2.

Finally, employing the relation between the absorption probability and
the greybody factor, Eq.(\ref{grey}), leads to 
\beq
\sigma_\ell(\om)=\frac{4 \pi^2\,(2\ell+1)}{(n+1)^2}\,
\frac{\Ga(\frac{\ell+1}{n+1})^2\,\Ga(1+\frac{\ell}{n+1})^2}
{\Ga(\frac{1}{2}+\ell)^2\,\Ga(1 +\frac{2\ell+1}{n+1})^2}\,
\biggl(\frac{\om r_H}{2}\biggl)^{2\ell}\,r_H^2\,.
\label{finalresult}
\eeq
In this equation, we have set $n=0$ in Eq.(\ref{grey}) as the
partial waves are purely confined to the 3-spatial-dimensional
brane, the only dependence on $n$ being in $|{\cal A}|^2$.

\begin{center}
$\begin{array}{|c||c||l|} \hline \hline 
\multicolumn{3}{|c|}{\rule[-3mm]{0mm}{8mm} 
{\bf Table~2: \rm \,\,Absorption\,\,probabilities\,\,for\,\,a\,\,
(4D)\,\,brane\,\,scalar\,\,field}} \\ \hline\hline
{\rule[-3mm]{0mm}{8mm}
\hspace*{0.8cm}{\bf n=2} \hspace*{0.8cm}} & \hspace*{0.8cm} \ell=0 \hspace*{0.8cm}
& \hspace*{0.8cm} 
|{\cal A}|^2 \simeq 4\,(\om r_H)^2 +\ldots \hspace*{0.8cm}\\[3mm]
& \ell=1 & \hspace*{0.8cm}|{\cal A}|^2 \simeq \frac{16 \pi^2}{243}\,(\om r_H)^4 
+ \ldots \hspace*{0.8cm}\\[3mm]
& \ell=2 & \hspace*{0.8cm}|{\cal A}|^2 \simeq \frac{4}{(15)^2}\,(\om r_H)^6 +
 \ldots\hspace*{0.8cm}\\[3mm] \hline
{\rule[-3mm]{0mm}{8mm}
\hspace*{0.8cm}{\bf n=4}} \hspace*{0.8cm} & \ell=0 &
\hspace*{0.8cm}|{\cal A}|^2 \simeq 4\,(\om r_H)^2 + \ldots
\hspace*{0.8cm}\\[3mm]
& \ell=1 & \hspace*{0.8cm}|{\cal A}|^2 \simeq \frac{4}{25}\,
\frac{\Ga(2/5)^2\,\Ga(6/5)^2}{\Ga(8/5)^2}\,(\om r_H)^4 + \ldots
\hspace*{0.8cm}\\[3mm]
& \ell=2 &
\hspace*{0.8cm}|{\cal A}|^2 \simeq \frac{4\,\Ga(3/5)^2\,\Ga(7/5)^2}{(15)^2}
\,(\om r_H)^6 + \ldots
\\[3mm] \hline
{\rule[-3mm]{0mm}{8mm}
\hspace*{0.8cm}{\bf n=6}} \hspace*{0.8cm} & \ell=0 &
\hspace*{0.8cm}|{\cal A}|^2 \simeq 4\,(\om r_H)^2 + \ldots
\\[3mm]
& \ell=1 &
\hspace*{0.8cm}|{\cal A}|^2 \simeq \frac{4}{49}\,
\frac{\Ga(2/7)^2\,\Ga(8/7)^2}{\Ga(10/7)^2}\,(\om r_H)^4 + \ldots
\\[3mm]
& \ell=2 &
\hspace*{0.8cm}|{\cal A}|^2 \simeq \frac{4}{(21)^2}\,
\frac{\Ga(3/7)^2\,\Ga(9/7)^2}{\Ga(12/7)^2}\,(\om r_H)^6 + \ldots
\\[3mm] \hline \hline

\end{array}$

\end{center}

Keeping $n$ fixed and varying $\ell$, 
we see that the absorption probability decreases once again as
$\ell$ increases (see Table 2): the dominant term becomes more and
more suppressed
by extra powers of $(\omega r_H)$ and its numerical coefficient also 
decreases (the introduction of the multiplicity of states with the 
same angular momentum number $\ell$ does not change this behaviour). 
If we fix instead $\ell$ and vary $n$, a radically different behaviour,
from the one observed in the case with a bulk scalar field, emerges: 
the leading term, in the expansion of $|{\cal A}|^2$ in powers
of $(\omega r_H)$, remains the same, since it is $n$-independent,
while its coefficient increases as $n$ increases. In other words, for
a given partial wave, the absorption probability, and therefore the
greybody factor, increases as the number of extra dimensions being
projected on the 3-brane also increases.

\section{Conclusions}

In this paper, we have studied the problem of scalar emission in a spherically
symmetric $D$-dimensional Schwarzschild black hole background. The cases
of the emission of a $(4+n)$-dimensional bulk scalar field and of a 4-dimensional
brane-localized scalar field were studied separately and the greybody factor
was determined in each case. This quantity causes
the spectrum of Hawking radiation to deviate from the black body spectrum as
it strongly depends on the energy of the particle mode emitted. Moreover,
it encodes information about the gravitational background and thus on the
number of extra dimensions that might exist (both in the case where the scalar 
field is free to propagate in the $(4+n)$-dimensional bulk or when the field `feels'
the existence of extra dimension only through the induced metric on our 3-brane).

The differential equation for the radial part of the scalar field 
can be solved by an approximation method valid in leading order in
$(\om r_H)$: the solution of this equation was found in the `near-horizon'
and `far-field' region and were subsequently `stretched' and matched in an
intermediate regime. This matching allows us to determine the absorption
coefficient for scattering in the black-hole background, which then leads
to the greybody factor, $\sigma(\om)$, according to Eq. (\ref{grey}). 

We first focused on the case of a bulk scalar field
propagating in a $(4+n)$-dimensional Schwarzschild black hole background.
The general form for the amplitude probability was determined and an
analytical, simplified version was also presented that allowed us to
display the leading functional dependence on the expansion parameter
$(\omega r_H)$ in terms of the number $n$ of extra dimension and the
angular momentum number $\ell$.  Our results in this case are presented in
Eq.(\ref{siresult}) and in Table~1.

The most phenomenologically interesting case is the one of a scalar
field that is confined on a 3-brane and propagates in the induced
spacetime background of a black hole (which is necessarily
higher-dimensional). The functional form 
of the resulting greybody factor depends only on the angular momentum
number, $\ell$, through $(\omega r_H)^{2\ell+2}$. 
The dependence on the number of extra dimensions is entirely contained within
the coefficient of this leading term.  Our primary results are given
in Eq.(\ref{finalresult}) and in Table~2.

It is tempting to compare the results derived in the case of
a brane-localized scalar field (which nevertheless is part of a
higher-dimensional manifold) with those valid in the case where
a purely 4-dimensional scalar field propagates in a Schwarzschild 
black-hole background.
Both cases lead to the same value of $|{\cal A}|^2$ for an $s$-wave
and, thus, no distinction can be made between the two backgrounds.
However, for higher partial waves, the value of the absorption 
probability in the case of a brane scalar field is always larger
than the one for a purely 4D field, a fact which in principle can
be used to distinguish between the two cases.

In a companion paper~\cite{promise} we employ the
techniques developed in the current
paper to derive the greybody factors for higher-spin fields localized on a
brane. This allows allows us to discuss the physics and phenomenology of
black hole decay, as might be observed at the LHC.


\section*{Appendix}

For completeness we here present some of the calculation of
scalar emission in the case
where the scalar field propagates in a purely 4-dimensional Schwarzschild
background without the assumption of the presence of extra dimensions.
The expression of the absorption coefficient $|{\cal A}|^2$ can be easily
found by first putting $n=0$ (the number of projected extra dimensions on
the 4-dimensional plane) in the result for the ratio $B_+/B_-$.
Then, Eq.(\ref{B-p}) becomes
\beq
\frac{B_+}{B_-} = -\biggl(\frac{2}{\om r_H}\biggl)^{2\ell+1}\,
\frac{\Ga(\ell +\frac{1}{2})^2\,\Bigl(\ell +\frac{1}{2}\Bigl)\,
\Ga(1-2\beta)\,\Ga(\alpha+\beta)^2}{\pi\,\Ga(1+\alpha-\beta)^2\,
\,\Ga(2\beta-1)}\,,
\eeq
where now
\beq
\alpha = -i \om r_H\,, \qquad
\beta =\frac{1}{2}\,\biggl(1 - \sqrt{(2l +1)^2 -4(\om r_H)^2}\biggl)\,.
\label{al-be-4}
\eeq
By using then Eq.(\ref{sigma}), we obtain the result for $|{\cal A}|^2$
which, in its simplified form, reads
\beq
|{\cal A}|^2= 16 \pi\,
\biggl(\frac{\om r_H}{2}\biggl)^{2\ell+2}\,
\frac{\Ga(\ell+1)^4}
{\Ga[\frac{1}{2}+\ell]^2\,\Ga[2\ell +2]^2}\,.
\eeq
Note that the same results follow
by putting $n=0$ in all the expressions of section 2.1, as expected.

We display the results for the absorption coefficient in the pure
4-dimensional case and for the values $\ell=0,1,2$ of the angular
momentum number in Table~3.

\begin{center}
$\begin{array}{|c||l|} \hline \hline 
\multicolumn{2}{|c|}{\rule[-3mm]{0mm}{8mm} 
{\bf Table~3: \rm \,\,Absorption\,\,probabilities\,\,for\,\,a\,\,
(4D)\,\,scalar\,\,field}} \\ \hline\hline
{\rule[-3mm]{0mm}{8mm}
\hspace*{1cm} \ell=0} \hspace*{1cm} & \hspace*{1.2cm} 
|{\cal A}|^2 \simeq 4\,(\om r_H)^2 +\ldots \hspace*{0.8cm}\\[3mm]
\ell=1 & \hspace*{1.2cm}|{\cal A}|^2\simeq \frac{1}{9}\,(\om r_H)^4 + \ldots
\\[3mm]
\ell=2 & \hspace*{1.2cm}|{\cal A}|^2\simeq \frac{1}{(45)^2}\,(\om r_H)^6 + \ldots
\\[3mm] \hline \hline
\end{array}$
\end{center}

\end{document}